\documentclass[%
 reprint,
superscriptaddress,
 amsmath,amssymb,
 aps,
prx,
floatfix,
]{revtex4-2}

\usepackage{graphicx}
\usepackage{dcolumn}
\usepackage{bm}
\usepackage{cancel}%
\usepackage{xcolor}
\usepackage[normalem]{ulem}

\newcommand{\RNum}[1]{\uppercase\expandafter{\romannumeral #1\relax}}

\begin{document}

\title{Energetic rigidity \RNum{2}.\\
Applications in examples of biological and underconstrained materials}%

\author{Ojan Khatib Damavandi}
\affiliation{%
 Department of Physics, Syracuse University, Syracuse, New York 13244, USA}%
\author{Varda F.\ Hagh}
\affiliation{%
 James Franck Institute, University of Chicago, Chicago, IL 60637, USA}%
\author{Christian D.\ Santangelo}
\email[To whom correspondence should be addressed:\\]{cdsantan@syr.edu}
\affiliation{%
 Department of Physics, Syracuse University, Syracuse, New York 13244, USA}
\author{M.\ Lisa Manning}
\email[To whom correspondence should be addressed:\\]{mmanning@syr.edu}
\affiliation{%
 Department of Physics, Syracuse University, Syracuse, New York 13244, USA}

\begin{abstract}
This is the second paper devoted to energetic rigidity, in which we apply our formalism to examples in two dimensions: underconstrained random regular spring networks, vertex models, and jammed packings of soft particles. Spring networks and vertex models are both highly underconstrained, and first-order constraint counting does not predict their rigidity, but second-order rigidity does. In contrast, spherical jammed packings are overconstrained and thus first-order rigid, meaning that constraint counting is equivalent to energetic rigidity as long as prestresses in the system are sufficiently small. Aspherical jammed packings on the other hand have been shown to be jammed at hypostaticity, which we use to argue for a modified constraint counting for systems that are energetically rigid at quartic order.
\end{abstract}


\maketitle

\section{\label{sec:intro} Introduction}

Biological materials exhibit remarkable mechanical responses. They are able to dynamically tune their mechanical properties -- such as their fluidity or elastic moduli -- in localized regions and in response to stimuli.
In particular, it is now clear that many organisms tune the mechanical properties of tissues across a fluid-to-solid transition to perform tasks, such as elongating a body axes during development~\cite{mongera2018fluid,Wang2020}, enhancing barrier function in maturing epithelia~\cite{Angelini2011, Park2015,devany2021cell}, or facilitating the escape of metastatic cancer cells~\cite{grosser2021cell,ilina2020cell}.

Some of these fluid-to-solid transitions are governed by a change in the cell density and free volume~\cite{petridou2021rigidity,kim2021embryonic,mongera2018fluid}, similar to the transitions in jammed sphere packings, and therefore can be understood by constraint counting arguments. If there are more constraints than degrees of freedom, the system is rigid. As discussed in the companion paper~\cite{damavandi2021a}, constraint counting is equivalent to considering whether first-order perturbations to the constraints are allowed by the geometry of the network.

In contrast, for confluent tissues -- where there are no gaps or overlaps between cells and so the packing fraction is always unity -- experiments~\cite{Angelini2011,Sadati2013,Kasza2014,Garcia2015,Park2015,Park2016,Wang2020} and computational models \cite{Farhadifar2007, Staple2010, Bi2015, Bi2016, Moshe2018,Merkel2018,Yan2019} indicate that the rigidity transition is strongly correlated with changes to cell shape. Similarly, experiments~\cite{Xu2000,Rammensee2007,Koenderink2009,Erk2010,Burla2019} and models~\cite{Tang1988, Storm2005, Wyart2008, Huisman2011,Sheinman2012, Feng2016, Sharma2016, Vermeulen2017,Rens2018,Merkel2019,Shivers2019} of biopolymer networks show that applied strains can rigidify the system. These rigidity transitions do not involve changes to constraints or network topology, and are driven instead by tuning a continuous control parameter -- the cell shape in biological tissues and applied strain in biopolymer networks. Indeed, both examples are highly underconstrained and thus constraint counting fails to predict their rigidity transition. 

So how and why does naive constraint counting fail? Seminal work by Yan and Bi~\cite{Yan2019} emphasizes that the rank of the full Hessian matrix, the matrix of second derivatives of the energy including effects beyond first-order perturbations, determines rigidity in vertex models for cellularized biological tissues. Significant strides in understanding the rigidity of underconstrained spring networks and vertex models of epithelial tissues have also been made by Merkel \textit{et al.}~\cite{Merkel2019} who suggested that rigidity transition in both types of systems can be understood in terms of a geometric incompatibility, where the local constraints on cells or fibers are incompatible with the global constraints imposed by the shape of the box. Additionally, they demonstrated that the rigidity transition coincides with the appearance of a system-spanning state of self stress. For finite-size systems, the state of self stress can be used to predict the scaling of elastic properties near transition, although other numerical results suggest that a different scaling may arise in the thermodynamic limit~\cite{Arzash2020}. Importantly, it has remained unclear how the geometric incompatibility leads to creation of such a single self stress that rigidifies these systems, even though they possess an extensive number of floppy modes. 

 In the companion paper~\cite{damavandi2021a}, we develop a formalism for energetic rigidity, whether a deformation raises the elastic energy of a structure, and relate it to other proxies for rigidity, including first-order and second-order rigidity.  Here, we analytically and numerically extend that formalism to apply to specific examples, including computational models for confluent tissues and fiber networks.  Unlike jammed packings of soft particles that are overconstrained and first-order rigid, we show the rigidity transition in vertex models and underconstrained spring networks is generated by second-order rigidity, meaning that perturbations to the constraints cost energy only to second order. 
 
 This suggests that underconstrained biological tissues near the rigidity transition are poised at a very special geometry: there is an extensive number of orthogonal deformations that cost zero energy to first order, and yet all the second-order perturbations are finite-cost. This result opens the door to a host of new questions important for the function of biological materials: what are the low-energy excitations in the rigid phase, likely excited in the presence of fluctuations? Is the linear vibrational spectrum sufficient to understand them? Does second-order rigidity generate universal features in the spatial structure of the states of self stress, and do organisms take advantage of such structure for patterning? In this work, we focus on characterizing features of the linear vibrational spectrum, and leave remaining questions for future work.
 
Also, in the particular case of unstressed underconstrained systems, the formalism results in a very simple counting argument for quadratic and quartic modes.  This counting argument explains features of the recent results on models for deformable, cell-like particles by Treado \textit{et al.}~\cite{Treado2021}, and is consistent with the previously observed vibrational structure seen in ellipsoids and other non-spherical packings~\cite{Donev2007, Mailman2009, Zeravcic2009, VanHecke2009,VanderWerf2018}.

\section{Overview of energetic rigidity and its proxies}\label{sec:formalism}

Similar to our companion paper~\cite{damavandi2021a}, we study systems with $N_{dof}$ generalized coordinates, $ \{x_n\}$, and $M$ constraints $f_\alpha( \{x_n\} )$ with energy function $E = \sum_{\alpha=1}^M f_\alpha^2/2$. The shear modulus is defined as the second derivative of the energy with respect to a shear variable, $\gamma$, in the limit of zero shear~\cite{Merkel2018,Wang2020}:
\begin{align}
    G &= \frac{1}{V}\frac{\mathrm{d}^2E}{\mathrm{d}\gamma^2}\nonumber\\
    &=\frac{1}{V}\left(\frac{\partial^2E}{\partial \gamma^2}-\sum_l\frac{1}{\lambda_l}\left[\sum_n\frac{\partial^2E}{\partial \gamma \partial x_n}u^{(l)}_n\right]\right),
    \label{eq:shearModulus}
\end{align}
where $V$ is the volume of the system, $\lambda_l$ and $u^{(l)}_n$ are the eigenvalues and eigenvectors of the Hessian matrix $H_{nm}=\partial^2 E/\partial x_n \partial x_m$ respectively, and the sum excludes eigenmodes with $\lambda_l=0$. The Hessian matrix can be written out in the following form,
\begin{align}
    H_{nm} &= \frac{\partial^2 E}{\partial x_n \partial x_m} = \sum_\alpha  \left[\frac{\partial f_\alpha}{\partial x_n}\frac{\partial f_\alpha}{\partial x_m} + f_\alpha\frac{\partial^2 f_\alpha}{\partial x_n\partial x_m} \right]\nonumber \\
    &=(R^T R)_{nm} + \mathcal{P}_{nm},
    \label{eq:Hessian}
\end{align}
where $R_{\alpha n } = \partial f_\alpha/\partial x_n$ is the rigidity matrix. We continue to use the terminology that is established in our companion paper~\cite{damavandi2021a}, where we refer to $(R^T R)_{nm}$ as the Gram term, and to $\mathcal{P}_{nm}$ as the prestress matrix. Excluding the trivial Euclidean modes, the necessary (but not sufficient) condition for floppiness is
\begin{align}
\sum_{nm} \mathcal{P}_{nm} \delta x_n \delta x_m &= -\sum_\alpha   \left(\sum_n\frac{\partial f_\alpha}{\partial x_n} \delta x_n\right)^2.
\label{eq:floppiness-condition}
\end{align}

A linear (first-order) zero mode (LZM) ,$\delta x_n^{(0)}$, is defined as a motion that preserves the constraints $f_\alpha$ to linear order:
\begin{equation}
    \sum_n \frac{\partial f_\alpha}{\partial x_n}  \delta x_n^{(0)} = \sum_n R_{\alpha n } { \delta x_n^{(0)}} = 0.
    \label{eq:ZM}
\end{equation}
Excluding the Euclidean motions, a nontrivial LZM is often called a floppy mode (FM)\cite{Lubensky2015}. A state of self stress is a vector, $\sigma_{\alpha}$, in the left nullspace of the rigidity matrix, $\sum_{\alpha} \sigma_{\alpha} R_{\alpha n} = 0$, that allows stresses on the contacts while keeping each particle in force balance. The number of linear zero modes, $N_0$, and states of self stress, $N_s$, are related through the Maxwell-Calladine constraint count, $N_{dof}-M = N_0 - N_s$~\cite{Calladine1978}.

A second-order zero mode is a LZM that preserves the constraints $f_\alpha$ to second order, and must satisfy
\begin{eqnarray}\label{eq:secondorder}
    \sum_\alpha \sum_{nm} \sigma_{\alpha, I} \frac{\partial^2 f_\alpha}{\partial x_n\partial x_m}  \delta x_n^{(0)} \delta x_m^{(0)} = 0,
\end{eqnarray}
for all states of self stress $\sigma_{\alpha,I}$. If the only LZMs that satisfy Eq.~(\ref{eq:secondorder}) are the trivial rigid motions, the system is called \emph{second-order rigid} \cite{Calladine1991,Connelly1996}.

\section{Examples}
\label{sec:Examples}
We now use the formalism of energetic rigidity to study two examples of underconstrained systems, 2D random regular spring networks and vertex models, from analytical and numerical perspectives and examine whether their linear zero modes make them floppy. We then briefly examine rigidity of jammed packings in the context of this new formalism.

\begin{figure}[htp]
    \centering
    \includegraphics[width=0.9\linewidth]{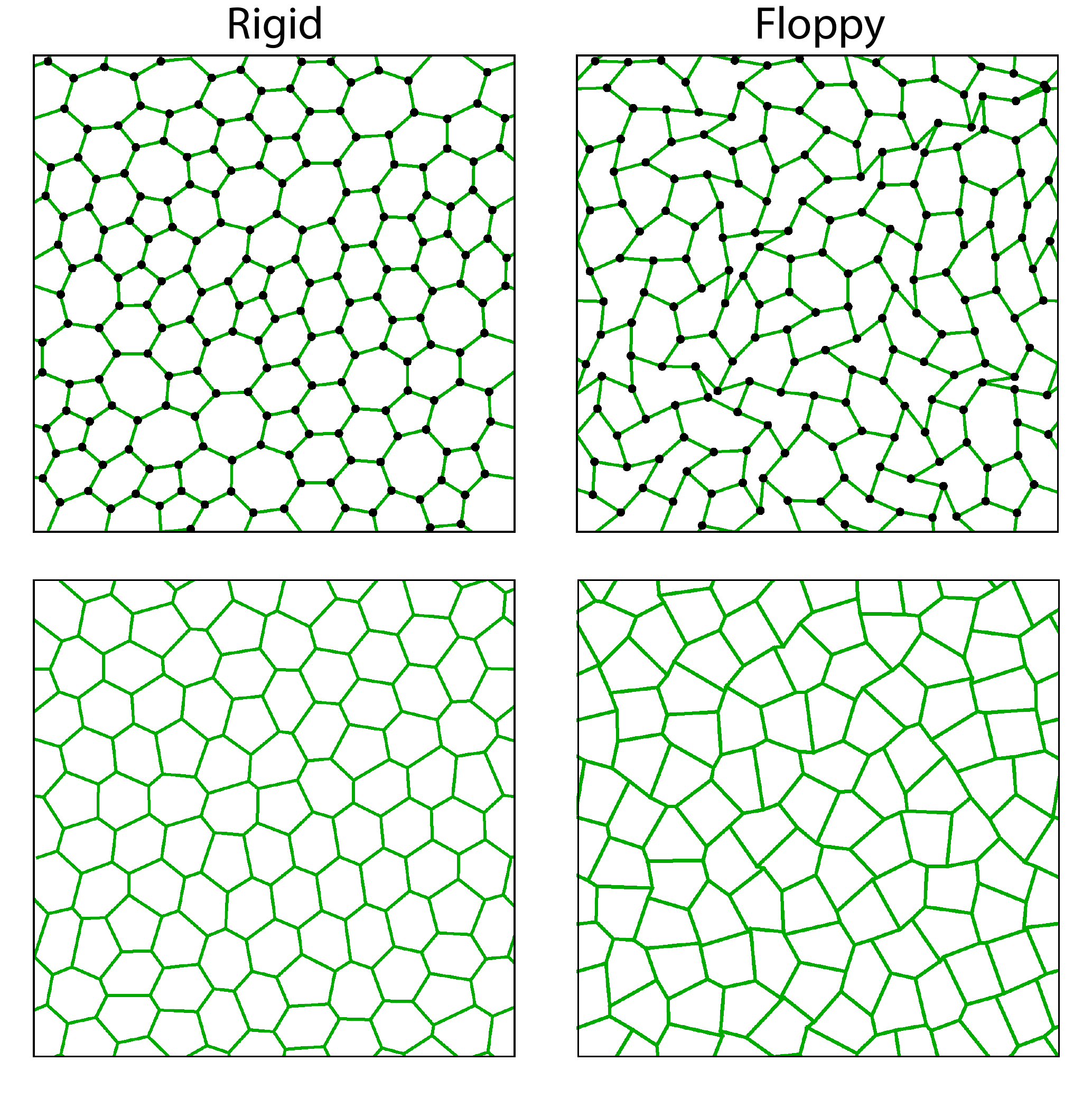}
    \caption{\textbf{Spring Network (Top) and Vertex Model (Bottom)} in rigid and floppy phases. Top left: A rigid spring network with $L_0 = 0.50$. Top right: A floppy spring network with $L_0 = 0.70$. Bottom left: A rigid vertex model with $P_0 = 3.50$. Bottom right: A floppy vertex model with $P_0 = 4.00$. The spring network and vertex model examples in this figure are prepared in a periodic box of area $A_{box} = 100$ with $N_{cell}=100$ cells and $N=200$ vertices.}
    \label{fig:spring-vertex}
\end{figure}

\subsection{2D Spring Network} \label{2D_Spring_Network}
Our first example is a 2D spring network comprised of $N$ vertices connected by springs with coordination number $z=3$ in a periodic square of fixed length (Fig.~\ref{fig:spring-vertex}). We choose $z=3$ so that the system is underconstrained and the network is regular, and there are no dangling vertices, but the results are valid for any $z<4$ (for $z > 4$ the system is overconstrained and first-order rigid). For simplicity, we assume the springs are identical with rest lengths $L_0$. For this system, $N_{dof} = 2N$ and $M = 3N/2$. From constraint counting, the number of LZMs, $N_0 \geq N/2$, so there are many system spanning FMs and one might be inclined to conclude that the system is floppy. However, we will see that for a range of spring rest lengths, the system possesses a self stress and becomes energetically rigid. 

\subsubsection{Analytical Results} \label{spring_network_analytical}
Calling the length of the $\alpha^{th}$ spring $L_\alpha$, the energy of this spring network is
\begin{equation}
    E = K_L \sum_{\alpha=1}^{3N/2} (L_\alpha - L_0)^2,
    \label{eq:springnet-energy}
\end{equation}
which defines the constraints $f_\alpha = L_\alpha - L_0$. Here, spring constants are identical and equal to $2K_L$. In Appendix~\ref{app:spring}, we show that the condition for FMs to be second-order (Eq.~(\ref{eq:secondorder})) can be written as
\begin{equation}
    \sum_\alpha \sigma_{\alpha,I}\frac{\left(\delta\mathbf{L}_\alpha \right)^2}{L_\alpha}=0. \label{eq:secondorder-spring}
\end{equation}
Assuming the edge $\alpha$ connects vertices $n$ and $m$, $\delta\mathbf{L}_\alpha = \delta \mathbf{X}_n^{(0)} - \delta\mathbf{X}_m^{(0)}$ is the vectorial displacement of the edge in response to FM $\{\delta \mathbf{X}_n^{(0)}\}$. Since trivial LZMs are excluded, this displacement must be perpendicular to the edge, $\delta\mathbf{L}_\alpha =\delta\mathbf{L}_\alpha^\perp$. An important observation is that if there exists a positive self stress ($\sigma_{\alpha,I}>0$), no FMs will satisfy Eq.~(\ref{eq:secondorder-spring}).

It can similarly be shown (Appendix~\ref{app:spring}) that the $G=0$ condition for spring networks is (Eq.~(\ref{eq:floppiness-condition}))
\begin{equation}
    \sum_\alpha (L_\alpha - L_0) \frac{\left(\delta\mathbf{L}_\alpha^\perp\right)^2}{2L_\alpha}= - \sum_\alpha \left( \delta L_\alpha^\parallel\right)^2,
    \label{eq:floppiness-condition-spring}
\end{equation}
for any global mode of motion, where $\delta {L}_\alpha^\parallel$ is the component of $\delta\mathbf{L}_\alpha$ parallel to the edge. The left hand side of Eq.~(\ref{eq:floppiness-condition-spring}) is equivalent to $\sum_{mn} \mathcal{P}_{mn} \delta x_n \delta x_m$ from the previous section.  

Next, using numerical models we show that even though the system has at least $N/2$ LZMs, there exists a positive self stress for $L_0 \leq L_0^*$ which implies energetic rigidity.

\subsubsection{Numerical Results}
We simulate the spring network by a random Voronoi tessellation of space in two dimensions which is guaranteed to produce networks with $z=3$. Details are given in Appendix~\ref{app:methods}. Defining the rigidity matrix as $\mathbf{R}_{\alpha m}=\partial f_\alpha/\partial \mathbf{X}_m$, where $\mathbf{X}_m$ is the position vector of vertex $m$, the LZMs are found by solving for the zero modes of $\mathbf{R}^T \mathbf{R}$. The states of self stress similarly are zero modes of $\mathbf{R} \mathbf{R}^T$. Fixing the box size, we vary $L_0$ and minimize the network, observing that the system goes through a transition from $f_\alpha = 0$ for $L_0 > L_0^*$ to $f_\alpha > 0$ for $L_0 < L_0^*$.

\begin{figure}[htp]
    \centering
    \includegraphics[width=0.8\linewidth]{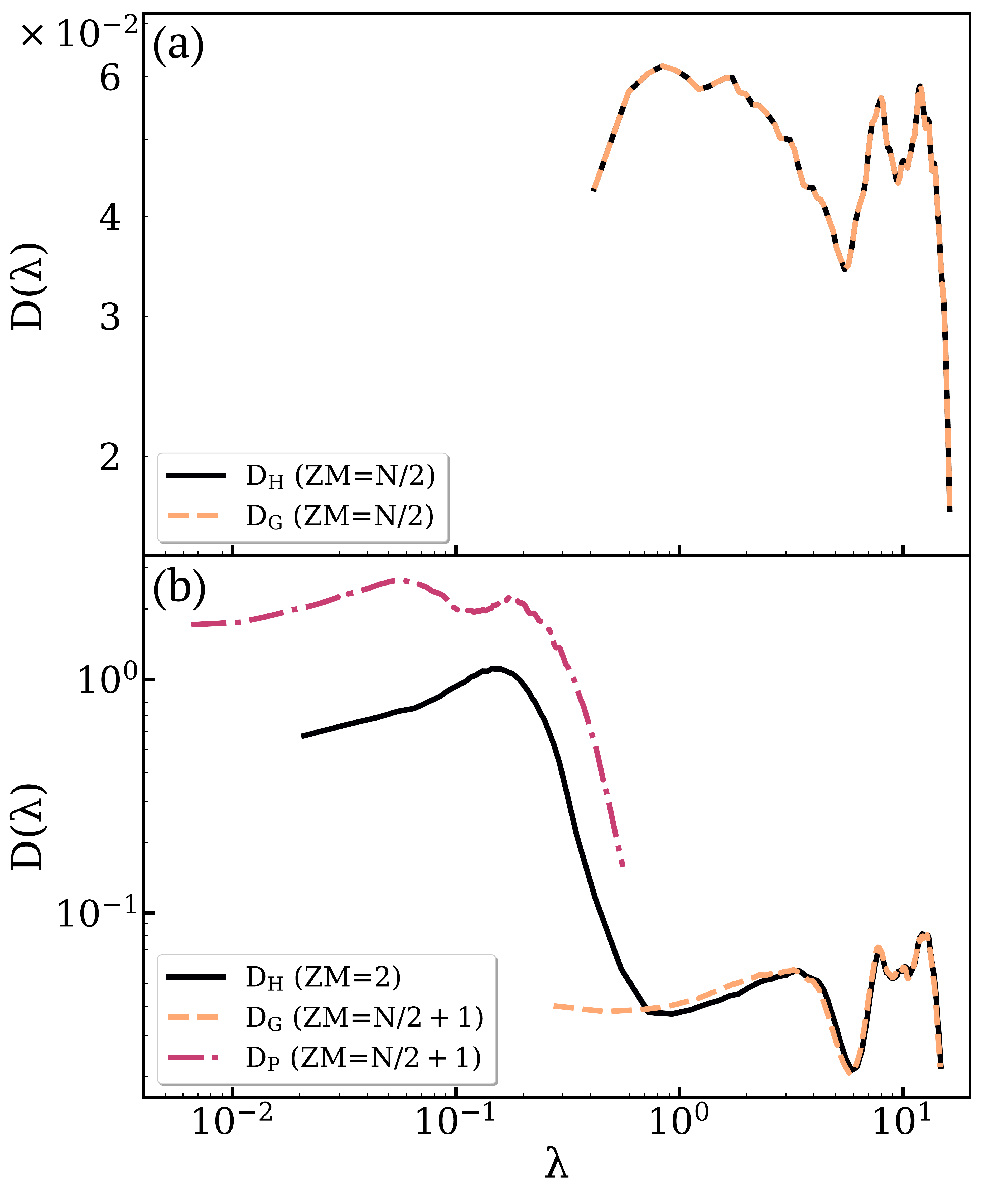}
    \caption{\textbf{Spring Network Density of States} $D$ as a function of the eigenvalue $\lambda$ for the Hessian matrix (solid black line), Gram matrix (dashed yellow line) and prestress matrix (dash-dotted red line), averaged over 10 samples with $N=1000$ and $K_L=2$. (a) In the Floppy regime ($L_0 = 0.65$), the Hessian and the Gram term both possess $N/2$ zero modes and their DOS curves overlap as there is no prestress in the network. (b) In the rigid regime ($L_0 = 0.61$), the Gram term possesses $N/2+1$ zero modes and $N_s =1$, while the Hessian only possesses two trivial (translational) zero modes. The Hessian DOS is dominated by the Gram term at high frequency and by the prestress matrix at low frequencies.}
    \label{fig:spring}
\end{figure}

From constraint counting, the number of LZMs $N_0$ minus the number of states of self stress $N_s$ is $N_0 - N_s = N/2$,
 which is confirmed numerically: we find $N_0 = N/2$ and $N_s = 0$ for $L_0 > L_0^*$, and $N_0=N/2+1$ and $N_s = 1$ for $L_0 < L_0^*$ (Fig.~\ref{fig:spring}). There are two trivial LZMs (no rotation is allowed because of periodic boundary conditions).  

Consistent with previous work~\cite{Merkel2018,Wang2020}, we use the shear modulus to quantify the energetic rigidity of the system. Specifically, we first calculate the eigenmodes of the Hessian, which then allows us to calculate the shear modulus $G$. Finding the eigenmodes of the Hessian and its components also allows us to plot the density of states (Figs.~\ref{fig:spring} and \ref{fig:spring_compare}b-d) as detailed in Appendix~\ref{app:methods}. From the observation that $f_\alpha = 0$ for $L_0 > L_0^*$ (Case 1  as defined in the companion paper~\cite{damavandi2021a}) and that there are many FMs, we would expect the system to be floppy in this regime.  We find that, indeed, $G=0$ for $L_0 > L_0^*$, agreeing with our prediction (Fig.~\ref{fig:spring_compare}a), and this is also consistent with previously reported results~\cite{Merkel2019}.

For $L_0 < L_0^*$, the Hessian has no nontrivial zero eigenmodes. We can see that by examining the eigenvalues of $\mathcal{P}_{mn}$ given by the LHS of Eq.~(\ref{eq:floppiness-condition-spring}), which is positive semi-definite since $f_\alpha > 0$. The high-frequency spectrum remains nearly identical to that in the unstressed case, which makes sense because the Gram term that dominates at high frequencies depends only on geometry, and does not change significantly as $L_0$ is lowered in the rigid regime. In contrast, in the rigid phase a new band of low-frequency modes appears in the Hessian, clearly coming from the prestress eigenspectrum. Moreover, the average magnitude of this low-frequency band is very sensitive to the control parameter $L_0$ -- eigenvalues shift to higher frequencies as $L_0$ is lowered (Fig.~\ref{fig:spring_compare}b).

\begin{figure}
    \includegraphics[width=0.8\linewidth]{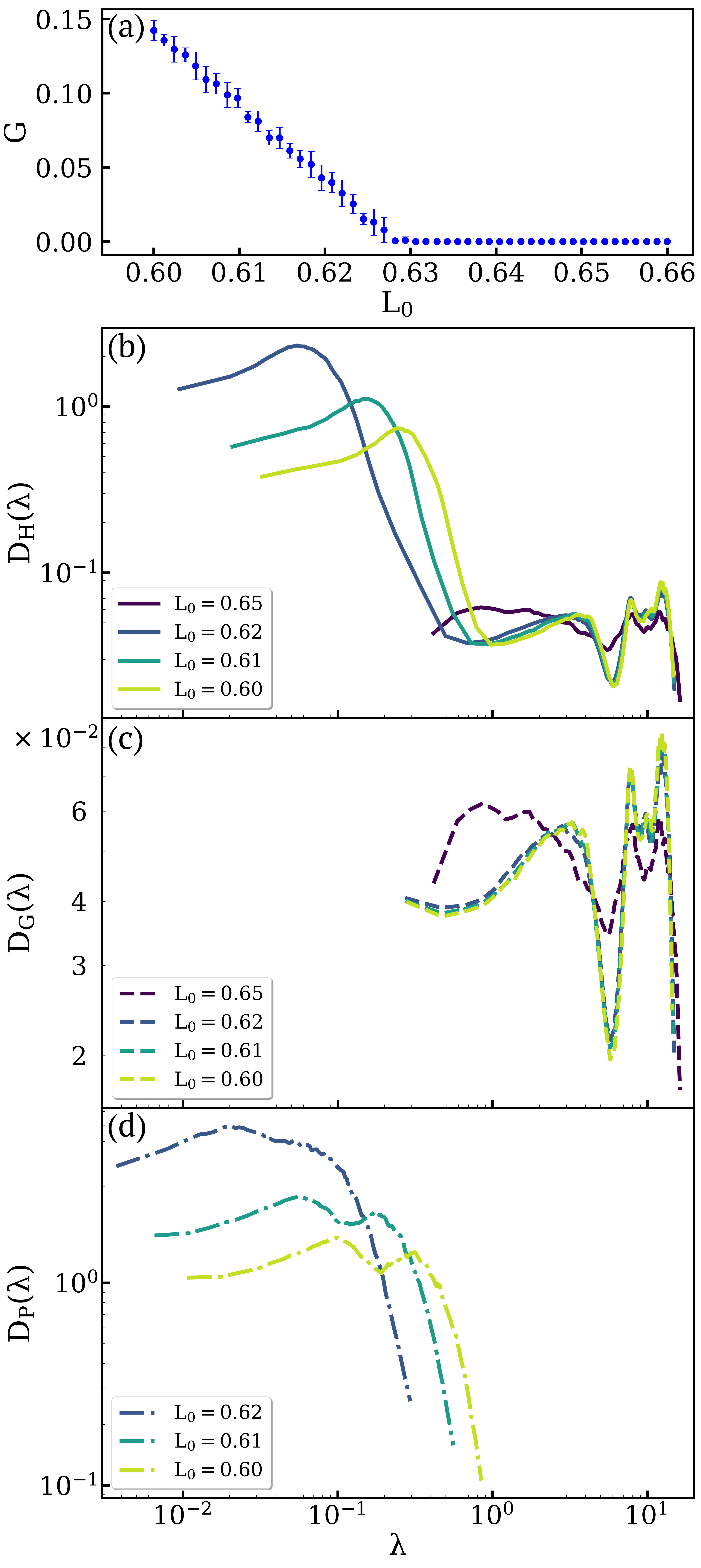}
    \caption{\textbf{Spring Network Comparison of Density of States} as a function of the eigenvalue $\lambda$ for the Hessian matrix ($D_H$), Gram matrix ($D_G$) and prestress matrix ($D_P$), averaged over 10 samples with $N=1000$ and $K_L=2$ for different values of $L_0$. (a) Shear modulus ($G$) of spring networks as a function of $L_0$ shows a transition from fluid to solid at $L_0^* \approx 0.63$. The error bars represent the standard deviation over 10 samples. (b) Transitioning from floppy to rigid coincides with the appearance of low frequency eigenmodes in the Hessian DOS that shift to higher frequencies as the system becomes stiffer. (c) Even though in the rigid regime there is only one new zero mode, the Gram matrix DOS looks significantly different when compared to systems with $L_0<L_0^*$ because the geometry of the system does not change significantly in the rigid regime. (d) Unlike the Gram matrix, the prestress increases linearly as $L_0$ is decreased in the rigid regime, as reflected by the shift in the prestress matrix DOS towards higher frequencies.}
    \label{fig:spring_compare}
\end{figure}

Numerical results indicate that the system only has one state of self stress, consistent with previous work~\cite{Merkel2019}. Therefore it falls under Case 2B -- second-order rigidity implies energetic rigidity -- and Eq.~(\ref{eq:floppiness-condition}) reduces to Eq.~(\ref{eq:secondorder-spring}) with $\sigma_{\alpha,I} \to f_\alpha$.
Due to the existence of the positive self stress, the system is second-order rigid and energetically rigid.
Exactly at the transition $L_0 = L_0^*$, the system is unstressed and $G=0$ (see \cite{Merkel2019}). However, at $L_0 \to L_0^{*-}$, since the system has a positive self stress and is second-order rigid, it is energetically rigid as well (Case 2A).
 
\subsection{2D Vertex Model}
Here we discuss rigidity of another highly underconstrained system, the 2D vertex model. The 2D vertex model consists of $N_{cell}$ polygonal cells tiling a 2D periodic square (Fig.~\ref{fig:spring-vertex}).  The energy of the system is 
\begin{equation}
    E =  \sum_{\alpha=1}^{N_{cell}}\left[ K_A (A_\alpha-A_0)^2 + K_P (P_\alpha-P_0)^2\right],
    \label{eq:vertexE}
\end{equation}
where $A_\alpha$ and $P_\alpha$ are the area and perimeter of the $\alpha^{th}$ cell, respectively. We have assumed that $A_\alpha$ has a preferred value $A_0$  and  $P_\alpha$ has a preferred value $P_0$. The energy is still Hookean but now is constructed from two sets of constraints $f_{\alpha,1} = A_\alpha - A_0$ and $f_{\alpha,2} = P_\alpha - P_0$. The total number of constraints is thus $M=2N_{cell}$. In the vertex model, DOFs are the vertices. Thus, in a periodic box, $N_{dof} = 4N_{cell}$ and we have at least $2N_{cell}$ LZMs from constraint counting.  These constraints are not all independent, however, because they act on the same vertices. In the numerical section, we will show by looking at the rank of the rigidity matrix that there is in fact one redundant constraint and there is a state of self stress and an extra zero mode because of it. We will also show that for $P_0<P_0^*$, a second state of self stress appears due to geometric incompatibility, similar to spring networks.

In the analytical results section, we limit ourselves to the more tractable version of the model with no area constraints so that $K_{A}=0$, and in the numerical section we will study the general version given by Eq.~(\ref{eq:vertexE}). 

\subsubsection{Analytical Results}\label{vertex_model_analytical}
We look the equations governing second-order zero modes and the $G=0$ condition (Eq.~(\ref{eq:floppiness-condition})) for the vertex model with no area term ($K_{A}=0$), thus $M=N_{cell}$. The constraints on the vertices are given by $f_\alpha = P_\alpha-P_0$.

The self stresses of $\mathbf{R}_{\alpha m}$ impose the following quadratic constraints on LZMs, $\delta \mathbf{X}_n$ (Appendix~\ref{app:vertex}):

\begin{align}
    \sum_{\alpha} \sigma_{\alpha,I} \sum_{\textrm{edge }j \ \in \textrm{ cell } \alpha} \frac{\left(\delta\mathbf{L}_j^\perp\right)^2}{L_j}=0.
    \label{eq:secondorder-vertex_2}
\end{align}

Since all vertices are connected to three edges and the box size is fixed, for a generic system, there are no nontrivial motions that do not introduce a $\delta\mathbf{L}_j^\perp$ and thus the inner sum is positive definite. Hence, similar to spring networks, if a self stress $\sigma_{\alpha,I}>0$ exists, the system is second-order rigid. To see if its shear modulus is zero, we again look at Eq.~(\ref{eq:floppiness-condition}):
\begin{align}
    \sum_{\alpha} f_\alpha &\sum_{\textrm{edge }j \ \in \textrm{ cell } \alpha}\frac{\left(\delta\mathbf{L}_j^\perp\right)^2}{L_j}\nonumber\\
    &= - \sum_\alpha \left( \sum_m \mathbf{R}_{\alpha m} \cdot \delta\mathbf{X}_m\right)^2.
    \label{eq:floppiness-condition-vertex}
\end{align}

We will see in the numerical results section and Appendix~\ref{app:vertex} that similar to spring networks, vertex model with $K_A=0$ has a positive state of self stress for $P_0 \leq P_0^*$ and thus is both second-order rigid and energetically rigid. For $P_0 > P_0^*$ however, all constraints are satisfied and the system has no state of self stress (Case 1). Therefore the system is floppy because it has many ($3N_{cell}$) LZMs. We will also see that vertex model with $K_A=1$ belongs to Case 2C when prestressed ($P_0 < P_0^*$), but it is still second-order rigid and energetically rigid.

\subsubsection{Numerical Results}
We simulate the vertex model (with both area and perimeter terms) using the same algorithm as spring networks but with the energy function given in Eq.~(\ref{eq:vertexE}) with $A_0 = 1$ and varying $P_0$ in a fixed periodic box ($A_{box}=N_{cell}$). Details, along with numerical results for systems with no area constraints ($K_A = 0$), are given in Appendix~\ref{app:vertex} and \ref{app:methods}. Each component of the rigidity matrix $\mathbf{R}_{\alpha m}$ now is a $2\times 2$ matrix: two components for the area and perimeter constraints associated with each cell $\alpha$ and two spatial components for each vertex $m$.

\begin{figure}
    \centering
    \includegraphics[width=0.8\linewidth]{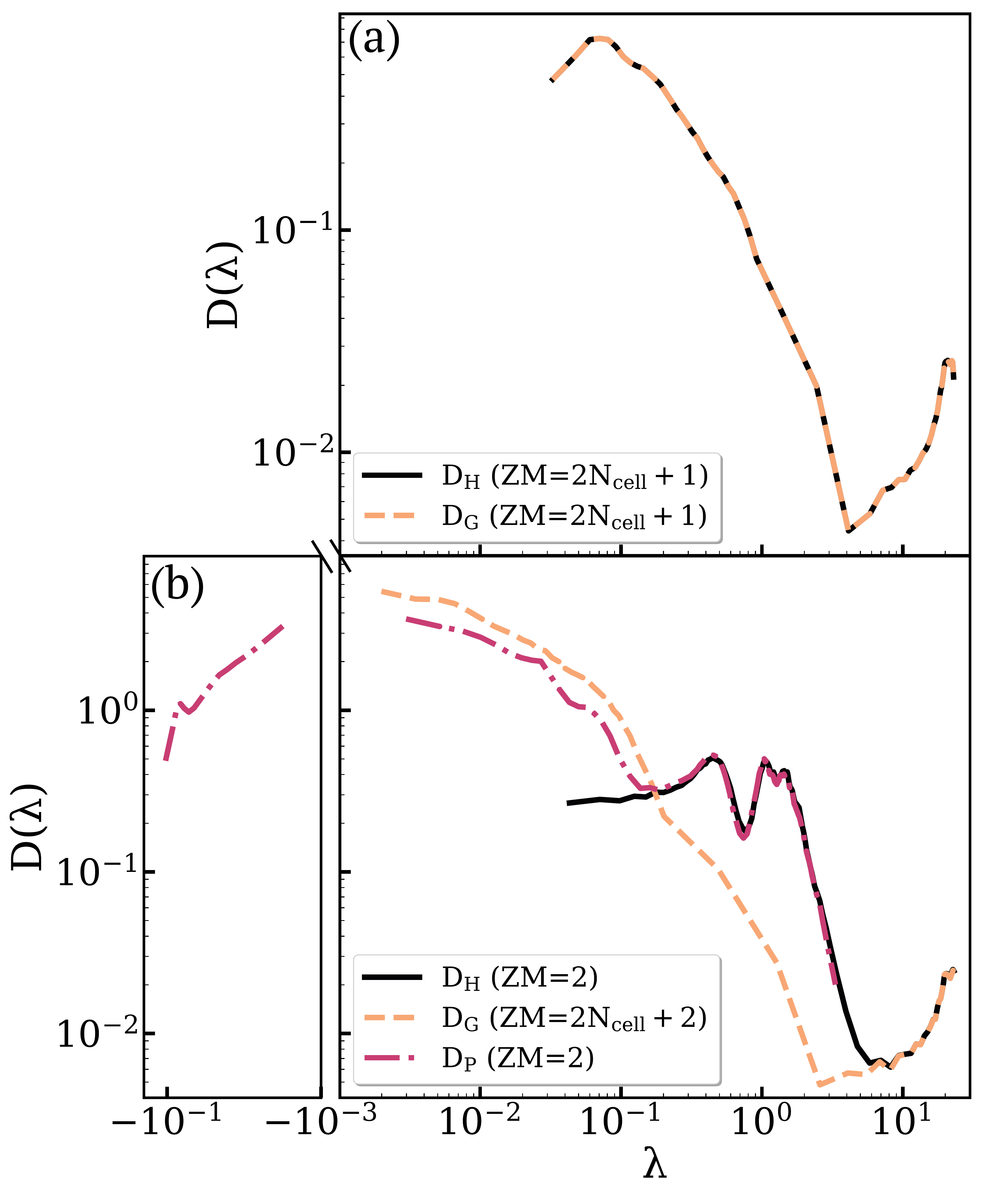}
    \caption{\textbf{Vertex Model Density of States} $D$ as a function of the eigenvalue $\lambda$ for the Hessian matrix (solid black line), Gram matrix (dashed yellow line) and prestress matrix (dash-dotted red line), averaged over 10 samples with $N_{cell}=500$ and $K_A=K_P=1$. (a) In the Floppy regime ($P_0 = 3.90$), the Hessian and the Gram term both possess $2N_{cell}+1$ zero modes and their DOS curves overlap as there is no prestress in the network. $N_s=1$ which is the trivial area self stress. (b) In the rigid regime ($P_0 = 3.70$), the Gram term possesses $2N_{cell}+2$ zero modes and $N_s =2$, while the Hessian only possesses two trivial (translational) zero modes. The Hessian DOS is dominated by the Gram term at high frequency and by the prestress matrix at low frequencies. While the prestress matrix has some negative eigenvalues, the vast majority are positive.}
    \label{fig:vertex}
\end{figure}

We find that for $P_0>P_0^*$, all constraints are satisfied and the Hessian zero eigenmodes are the same as the LZMs ($N_0 = 2N_{cell}+1$) (Fig.~\ref{fig:vertex}a). Even though there are no prestresses, we find $N_s =1$ (satisfying the rank-nullity theorem). To understand the source of this self stress, note that the sum of all the cell areas must be equal to the fixed box size. Thus $A_0$ merely changes the overall pressure of the system~\cite{Yang2017} and we can rewrite the energy function as:
\begin{align}
    E =  \sum_{\alpha=1}^{N_{cell}}\left[ K_A (A_\alpha-\frac{A_{box}}{N_{cell}})^2 + K_P (P_\alpha-P_0)^2\right]\nonumber\\
    +K_{A}N_{cell}(\frac{A_{box}}{N_{cell}}-A_0)^2. 
\end{align}
Therefore, changing $A_0$ amounts to increasing the overall tissue pressure without breaking force balance and by definition there must be a self stress associated with it. Indeed, we see that the area components of self stress are 1 while its perimeter components are 0: $\sigma = (1,0,\dots,1,0)$. This also means that one of the area constraints is redundant: if $N_{cell} -1$ cells have satisfied their constraints, the last cell automatically does as well.

\begin{figure}
    \includegraphics[width=0.8\linewidth]{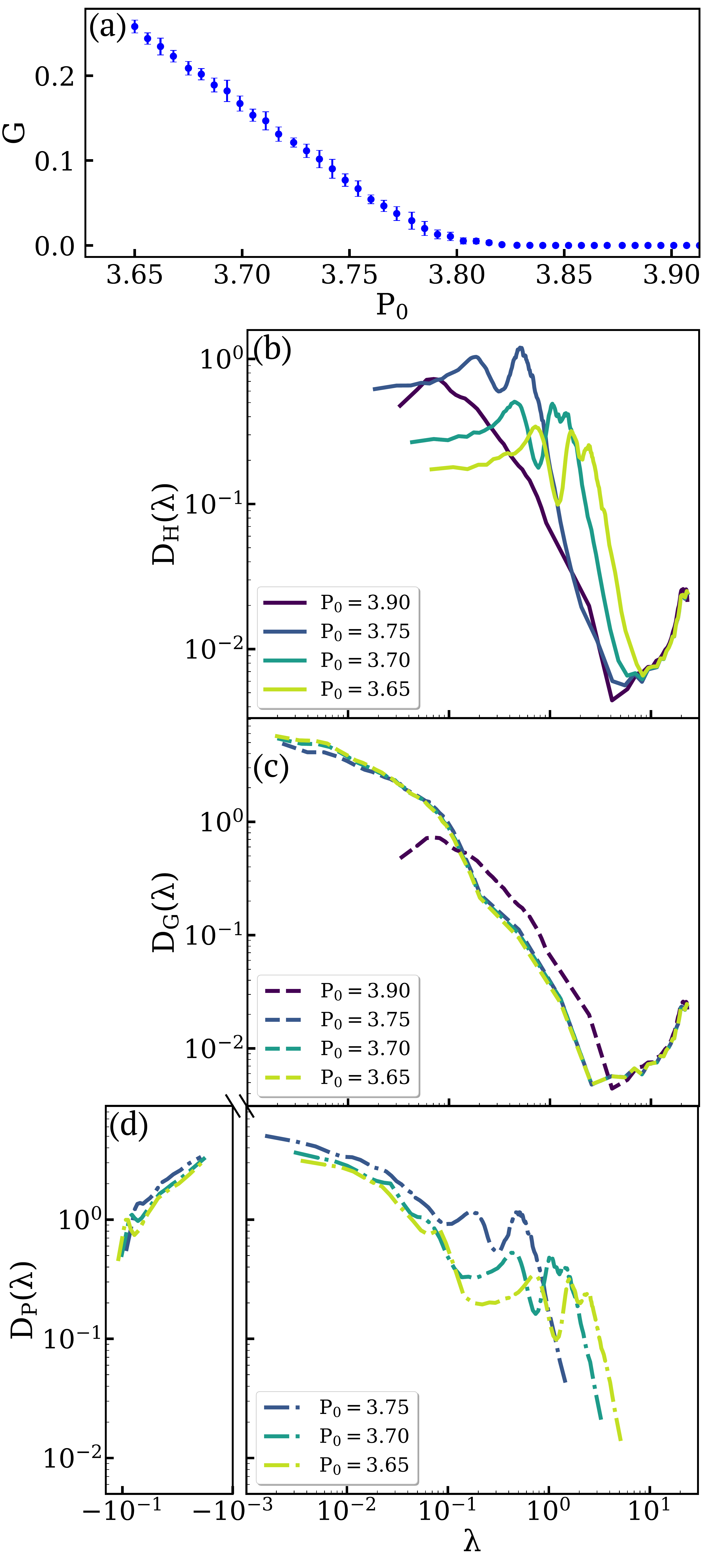}
    \caption{\textbf{Vertex Model Comparison of Density of States} as a function of the eigenvalue $\lambda$ for the Hessian matrix ($D_H$), Gram matrix ($D_G$) and prestress matrix ($D_P$), averaged over 10 samples with $N_{cell}=500$ and $K_A=K_P=1$ for different values of $P_0$. (a) Shear modulus ($G$) of vertex model with as a function of $P_0$ shows a transition from fluid to solid at $P_0^* \approx 3.84$. The error bars represent standard deviation over 10 samples. (b-d) Comparison between DOS of Hessian and its components for different values of $P_0$. (b) Transitioning from floppy to rigid coincides with the appearance of low frequency eigenmodes in the Hessian DOS that shift to higher frequencies as the system becomes more rigid. (c) Even though in the rigid regime there is only one new zero mode, the Gram matrix DOS looks significantly different when compared to systems with $P_0<P_0^*$ because the geometry of the system does not change significantly in the rigid regime. (d) Unlike the Gram matrix, the prestress increases linearly as $L_0$ is decreased in the rigid regime, as reflected by the shift in the prestress matrix DOS towards higher frequencies (both positive and negative).}
    \label{fig:vertex_compare}
\end{figure}

For $P_0<P_0^*$, none of the constraints are satisfied due to geometric incompatibility (Fig.~\ref{fig:vertex}b). The perimeter prestresses are all positive ($f_{\alpha,2}>0$), but the area prestresses are not. For $A_0 = 1$ and $A_{box} = N_{cell}$, the average, $\bar{f}_{\alpha,1} = 0$. We find $N_s = 2$ and $N_0 = 2N_{cell} + 2$. One of the self stresses is due to the geometric incompatibility we already saw in the spring network, $\sigma_1 = (A_1-A_{box}/N_{cell},P_1-P_0,\dots,A_{N_{cell}}-A_{box}/N_{cell},P_{N_{cell}}-P_0)$. The other one arises from the area constraint redundancy, $\sigma_2 = (1,0,\dots,1,0)$, and does not play an important role. We find $G=0$ for $P_0>P_0^*$, and $G>0$ for $P_0<P_0^*$ (Fig.~\ref{fig:vertex_compare}a) suggesting that the system is second-order rigid when $\sigma_2$ appears. Decreasing $P_0$ further shifts the Hessian eigenmodes to higher frequency and increases $G$ similar to spring networks (Fig.~\ref{fig:vertex_compare}b). This is again due to a shift in $\mathcal{P}_{nm}$ and not the Gram term (Figs.~\ref{fig:vertex_compare}c, d).

Looking at the spectrum of $\mathcal{P}_{nm}$, we can see that it has negative eigenvalues (Fig.~\ref{fig:vertex_compare}d) and thus the system falls under Case 2C. It is empirically still rigid because $\mathcal{P}_{nm}$ does not have directions negative enough to satisfy Eq.~(\ref{eq:floppiness-condition}). Even though its negative eigenvalues become more negative as $P_0$ is decreased, its positive ones dominate (Fig.~\ref{fig:vertex_compare}c). 

One way to identify whether it is the Gram or prestress matrix that is responsible for rigidity is to multiply the prestress matrix $\mathcal{P}_{nm}$ by an arbitrary $\epsilon > 1$. This suggests that the Hessian is dominated by the positive eigenvalues of the prestress matrix and not from a competition with the Gram term. For $K_A=0$ with $P_0<P_0^*$, $\mathcal{P}_{nm}$ is positive semi-definite and $N_s=1$, so it falls under Case 2B similar to spring networks. At the onset of rigidity, $P_0 \to P_0^{*-}$, both vertex models will show $G\to 0$, but since they both have a nontrivial self stress (Case 2A) and are second-order rigid, our formalism indicates they are energetically rigid as well.

\subsection{2D Jammed Packings}
Athermal packings of soft or hard spheres are a useful model for studying granular matter and glasses at zero temperature. A 2D disk packing with one state of self stress is in a way very similar to the spring networks that we studied above. States of self stress in jammed systems are extended over the entire system with compressive forces everywhere~\cite{Ellenbroek2015} which resembles the case of spring networks under tension. Here, the energy is given by:

\begin{equation}
    E = \frac{k}{2} \sum_{\alpha=1}^{2N - 1} h_{\alpha}^2 \ \Theta(h_{\alpha}),
    \label{eq:jamming-energy}
\end{equation}
where $h_{\alpha} = (1 - \frac{\rho_{\alpha}}{\sigma_{\alpha}})$ is the dimensionless overlap between particle pair $\alpha \equiv i, j$, with $\rho_{\alpha}$ being the distance between two disk centers and $\sigma_{\alpha}$ being the sum of their radii. The Heaviside step function is used to count contributions from positive overlaps only. 
\subsubsection{Analytical Results}
The analysis presented in section \ref{2D_Spring_Network} can also be used to describe the energetic rigidity in jammed packings of soft harmonic disks/spheres. One difference between a spring network under tension and a critically jammed packing under compression (with one state of self stress only) is that all of the prestress forces in a jammed packing are negative and therefore any terms in the expansion of the energy that are proportional to the first derivatives will be negative. Jammed packings thus belong to Case 2C. We do not present the analytical work for these systems here because it will be a repetition of the calculations in section \ref{spring_network_analytical}.

\subsubsection{Numerical Results}
\begin{figure}
    \centering
    \includegraphics[width=0.8\linewidth]{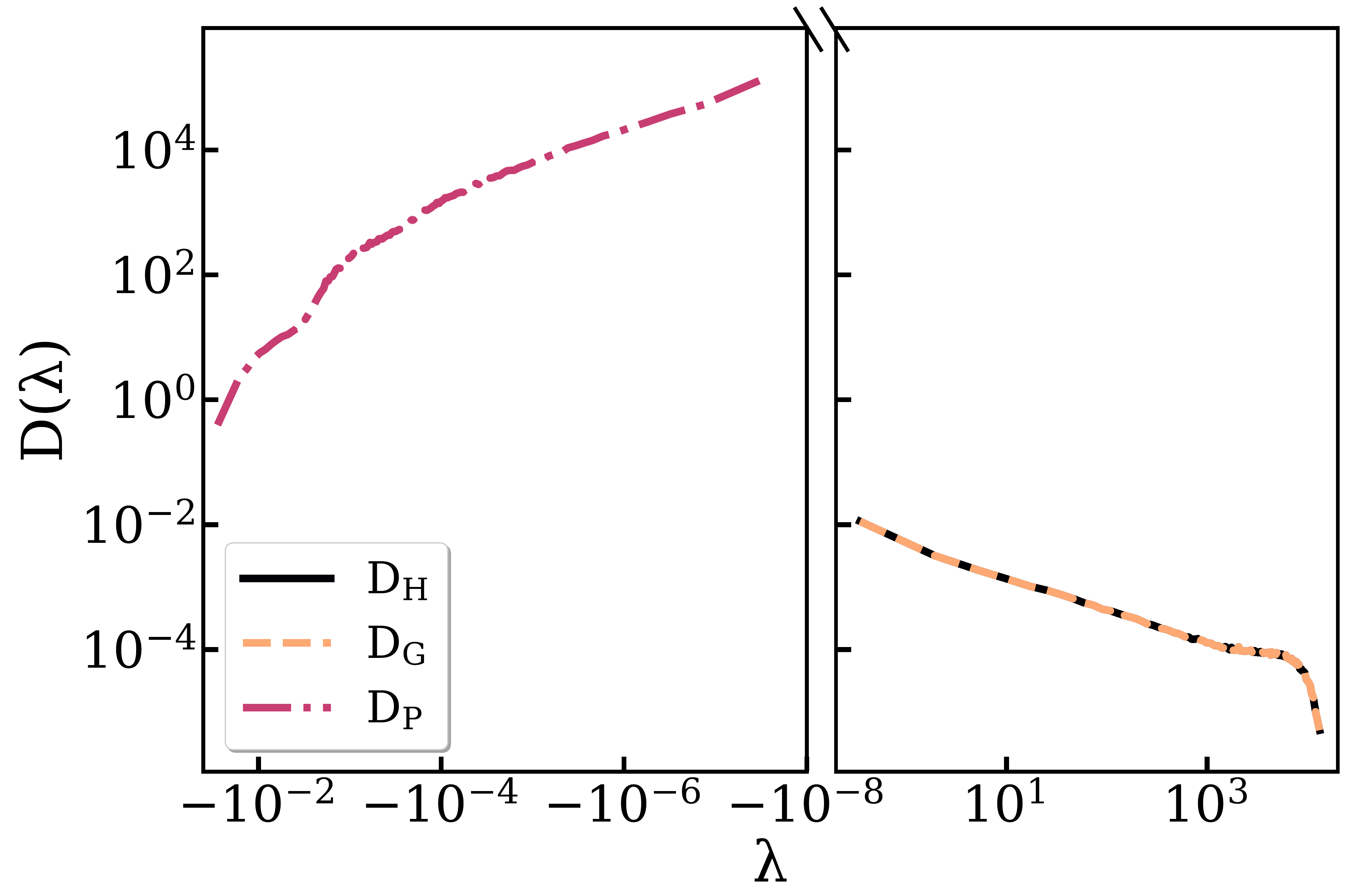}
    \caption{\textbf{Jammed packing Density of States} $D$ as a function of the eigenvalue $\lambda$ for the Hessian matrix (solid black line), Gram matrix (dashed yellow line) and prestress matrix (dash-dotted red line), in the rigid regime (at one state of self stress) averaged over 10 samples with $N=1000$. Each of these three data sets only have two zero modes per sample.}
    \label{fig:jamming}
\end{figure}

We create an ensemble of ten 2D disk packings very close to the critical jamming using standard methods as described in Appendix~\ref{app:methods}, so that there is one state of self stress in each packing. We calculate the eigenspectra of the Hessian as well as the Gram and prestress terms; the results are presented in Fig.~\ref{fig:jamming}.  

Unlike spring networks that have many non-zero modes in their floppy regime, a critically jammed packing will completely unjam and reach a global minimum with zero energy and zero eigenmodes everywhere if the density is lowered. Therefore, we do not report data from the floppy side of the transition. As can be seen from Fig.~\ref{fig:jamming}, the density of states of the full Hessian almost matches the density of states of the Gram term in Hessian. This is because at critical jamming, the prestress forces are infinitesimal and the magnitude of the eigenvalues of prestress term are several orders of magnitude smaller than their equivalent eigenvalues in the Gram term. At this point, the rigidity of the system is mainly determined by the Gram term in the Hessian. Since both Gram and prestress terms have two zero eigenmodes, the full Hessian also has two zero eigenmodes which is typical of a rigid body in 2D. 

A consequence of this is that the energetic rigidity of jammed systems \textit{can} be fully described using the Maxwell-Calladine count, since even at the jamming point where the pressure is zero and the prestress forces are infinitesimal, the system is first-order rigid. The prestress forces can only play a role in the energetic rigidity of the system when the pressure is large enough to push the system to an instability. Indeed it has been analytically shown that compressive prestresses can lower the shear modulus in amorphous solids~\cite{Cui2019}. This marks another difference between the spring networks under tension and soft harmonic particles under compression at one state of self stress. 

\subsection{Extended constraint counting in unstressed and weakly prestressed systems}

It is clear from the energetic rigidity formalism that, in general, constraint counting cannot always be generalized to explain the behavior of second-order rigid systems.  Nevertheless, there is one specific case, namely systems that are underconstrained, unstressed, and second-order rigid, where our formalism predicts a simple extended constraint counting.

Interestingly, this case encompasses both aspherical particles such as ellipsoids and may also include deformable particles. Specifically, over a decade ago it was shown that aspherical particles can comprise stable packings while underconstrained even in the limit of infinitesimal asphericity~\cite{Donev2007, Mailman2009, Zeravcic2009, VanHecke2009}. Such packings exhibit quartic modes of excitation~\cite{Mailman2009,VanderWerf2018}, and are stabilized because finite rotations are blocked by the curvature of particles at the contacts. Moreover, Donev and collaborators have explicitly shown that such packings become second-order rigid at jamming onset when they are unstressed~\cite{Donev2007}. 

For packings that are second-order rigid and unstressed, our formalism confirms they must also be energetically rigid at quartic order. Thus, any of the $N_{dof}$ directions at the energy minimum that are flat to quadratic order must increase at quartic order. Let $n_{quad}$ be the number of quadratic excitation modes, which is equal to the number of finite Hessian eigenvalues. Then the following equation must hold:
\begin{equation}
    N_{dof} = n_{quad} + n_{quart},
    \label{eq:modified_CC}
\end{equation}

\noindent
where $n_{quart}$ are the remaining excitation directions that must increase in energy at quartic order. We note that (~\ref{eq:modified_CC}) also applies to prestressed systems as long as second-order rigidity still implies energetic rigidity, i.e. $\mathcal{P}_{nm}$ is either positive semi-definite or small enough that does not destabilize the Hessian.

This rather trivial observation post-dicts constraint counting that has already been observed in friction-less ellipsoid packings~\cite{Mailman2009,VanderWerf2018}, which are unstressed and second-order rigid. Perhaps more usefully, it also predicts that a similar constraint counting should be valid in packings of deformable particles where the prestress matrix is small, and which are likely second-order rigid (although that remains to be confirmed). 

\section{Discussion and Conclusions}

In this paper, we demonstrate that the rigidity transition in many biological materials including cellularized confluent tissues and biopolymer networks is generated by second-order rigidity, and is not consistent with naive constraint counting. Instead, at the transition point these systems possess an extensive number of modes that are floppy to first order in the constraints, and yet all second-order perturbations to the constraints cost finite energy.

Our companion paper demonstrates that an important consideration is whether the prestress matrix is positive semi-definite or not in the rigid phase~\cite{damavandi2021a}. Therefore, we first focus on two examples where the prestress matrix is positive semi-definite: underconstrained spring networks and vertex models without an area constraint. These networks undergo a second-order rigidity transition with an emergent state of self stress. In both cases, we can analytically show that second-order rigidity implies energetic rigidity (\textit{i.e.} any small deformations in the system have an energy cost) both at the transition point where the shear modulus is zero and in the rigid regime, where the shear modulus is finite.  

In other examples where the prestress matrix is indefinite or negative semi-definite, we can still show analytically that second-order rigidity implies energetic rigidity at the transition point. However, away from the transition point, neither first-order nor second-order rigidity guarantee energetic rigidity. Moreover, we have identified two widely divergent examples in this category: vertex models with an area term and jammed spheres. Although we do not yet have analytic predictions for these systems away from the transition point, our numerical simulations indicate that in vertex models with an area term, second-order rigidity always implies energetic rigidity, while in jammed packings of soft spheres, first-order rigidity is sufficient to predict the onset of energetic rigidity. Interestingly, for prestressed vertex models, additional ``no-rotation'' constraints on vertices can be derived from the prestress matrix that explain the inability of vertices in vertex models to move when prestressed~\cite{liu2021}. These no-rotation constraints appear to be equivalent to second-order rigidity of vertex models.

These observation immediately give rise to an open question: is there a way to subdivide materials with indefinite or negative semi-definite prestress into two or more categories so that we can analytically predict which will be first-order or second-order rigid? One hint is that the susceptibility of the prestress matrix to additional pressure is quite different between our two examples; in jammed spheres the prestress matrix becomes more strongly negative definite when increasing the magnitude of the prestress along the direction of the self stress, while for vertex models with an area term the positive eigenvalues of the prestress matrix always remain dominant.

Another interesting question that remains unanswered involves the number of states of self stress that emerge in these networks when they undergo a second-order rigidity transition. Unlike first-order jammed systems where increasing the pressure leads to a quadratic increase in the number of contacts in excess of isostaticity (thereby the number of states of self stress)~\cite{parisi2020theory}, the second-order rigid examples discussed here seem to develop and maintain only one state of self stress in the rigid regime. How does this state of self stress evolve as a function of bond density and distance to the critical point?  What is its spatial structure, and how is that related to emergent geometric features such as fiber alignment~\cite{Kang2009}?

\begin{acknowledgments}
 We are grateful to Z. Rocklin for an inspiring initial conversation pointing out the connection between rigidity and origami, and to M. Holmes-Cerfon for substantial comments on the manuscript. This work is partially supported by grants from the Simons Foundation No 348126 to Sid Nagel (VH), No 454947 to MLM (OKD and MLM) and No 446222 (MLM). CDS acknowledges funding from the NSF through grant DMR-1822638, and MLM acknowledges support from NSF-DMR-1951921.
\end{acknowledgments}

\appendix

\section{Analytical calculations for spring networks}\label{app:spring}
Here we provide the details of our spring network calculations discussed in Section \ref{spring_network_analytical}.

It is useful to express $f_\alpha$ explicitly in terms of the DOFs, \textit{i.e.} the vertex positions $\mathbf{X}_n$ (note that here we are using a vectorial notation so $n \in \{1,\dots,N\})$. To do so, we define the incidence matrix
of the network, $I_{\alpha n}$

\begin{equation}
I_{\alpha n} = \left\{ \begin{array}{ccl}
1, & & \alpha \textrm{ leaves vertex } n\\
-1, & & \alpha \textrm{ enters vertex } n\\
0. & & \textrm{otherwise}
\end{array}\right.
\end{equation}
With this definition, we can rewrite the spring length as 
\begin{equation}
    L_\alpha = \Big\vert \sum_n I_{\alpha n}\mathbf{X}_n \Big\vert = \left[\sum_{n,m} I_{\alpha n}I_{\alpha m}\mathbf{X}_n \cdot \mathbf{X}_m\right]^{1/2}.
    \label{springL}
\end{equation}
and the constraints as
\begin{equation}
    f_\alpha = \left[\sum_{n,m} I_{\alpha n}I_{\alpha m}\mathbf{X}_n \cdot \mathbf{X}_m\right]^{1/2} - L_0
    \label{spring-f}
\end{equation}

Now we study behavior of the zero modes. To do that, we first perturb the network $\mathbf{X}_n \to \mathbf{X}_n+\delta\mathbf{X}_n$. Taylor expanding Eq.~(\ref{spring-f}) to second order in $\delta\mathbf{X}_n$, we find
\begin{align}
    &\delta f_\alpha = \frac{\left(\sum_n I_{\alpha n}\mathbf{X}_n  \right)\cdot\left(\sum_m I_{\alpha m}\delta\mathbf{X}_m \right)}{L_\alpha}+\nonumber\\ 
    &\frac{\left(\sum_n I_{\alpha n}\delta\mathbf{X}_n\right)^2}{2L_\alpha} -\frac{\left[\left(\sum_n I_{\alpha n}\mathbf{X}_n  \right)\cdot\left(\sum_m I_{\alpha m}\delta\mathbf{X}_m \right)\right]^2}{2L_\alpha^3}.
    \label{eq:deltaf_spring}
\end{align}
By comparing with Eq.~($8$) in the companion paper~\cite{damavandi2021a}, we determine the rigidity matrix of the system is defined as
\begin{equation}
    \mathbf{R}_{\alpha m} = \frac{\sum_n I_{\alpha n} I_{\alpha m} \mathbf{X}_n}{L_\alpha}.
    \label{R}
\end{equation}
We can then simplify Eq.~(\ref{eq:deltaf_spring}) 
\begin{align}
   &\delta f_\alpha =  \sum_m \mathbf{R}_{\alpha m} \cdot \delta\mathbf{X}_m \nonumber\\ 
    &+ \frac{\left(\sum_n I_{\alpha n}\delta\mathbf{X}_n\right)^2 - \left(\sum_m \mathbf{R}_{\alpha m} \cdot \delta\mathbf{X}_m\right)^2}{2L_\alpha}.
\end{align}

Linear zero modes are the solutions to $\sum_m \mathbf{R}_{\alpha m} \cdot \delta\mathbf{X}_m^{(0)}=0$. If a linear zero mode is also a second-order zero mode, it must additionally satisfy (Eq.~(\ref{eq:secondorder}))
\begin{equation}
    \sum_\alpha \sigma_{\alpha,I}\frac{\left(\sum_n I_{\alpha n}\delta\mathbf{X}_n^{(0)}\right)^2}{L_\alpha}=0. \label{eq:secondorder-spring-app}
\end{equation}
$\left(\sum_n I_{\alpha n}\delta\mathbf{X}_n\right)^2>0$ for any nontrivial motion, therefore if a positive state of self stress ($\sigma_{\alpha,I}>0$) exists, no zero mode will satisfy Eq.~(\ref{eq:secondorder-spring}). The $G=0$ condition is (Eq.~(\ref{eq:floppiness-condition})),
\begin{align}
    &\sum_\alpha f_\alpha \frac{\left(\sum_n I_{\alpha n}\delta\mathbf{X}_n\right)^2 - \left(\sum_m \mathbf{R}_{\alpha m} \cdot \delta\mathbf{X}_m\right)^2}{2L_\alpha} \nonumber\\
    &= - \sum_\alpha \left( \sum_m \mathbf{R}_{\alpha m} \cdot \delta\mathbf{X}_m\right)^2
    \label{eq:floppiness-condition-spring-app}
\end{align}
for any mode $\delta\mathbf{X}_n$. To retrieve Eqs.~(\ref{eq:secondorder-spring}) and (\ref{eq:floppiness-condition-spring}), we define  $\mathbf{L}_\alpha = \sum_n I_{\alpha n}\mathbf{X}_n$ to be the vector along the edge $\alpha$, and $\delta \mathbf{L}_\alpha = \sum_n I_{\alpha n}\delta\mathbf{X}_n$ to be its change due to perturbation $\delta \mathbf{X}_n$. The component of $\delta \mathbf{L}_\alpha$ parallel to $\mathbf{L}_\alpha$ is $\delta L_\alpha^\parallel = \delta L_\alpha = \sum_m \mathbf{R}_{\alpha m} \cdot \delta\mathbf{X}_m + \mathcal{O}(\delta \mathbf{X}_n^2)$.

\section{Analytic calculations for Vertex models}\label{app:vertex}
\subsection{Second-order rigidity of vertex model with $K_A=0$}
Here, we look the equations governing second-order zero modes and the $G=0$ condition for the vertex model with no area term ($K_{A}=0$) discussed in section \ref{vertex_model_analytical}, thus $M=N_{cell}$. The constraints on the vertices are given by
\begin{align}
    f_\alpha = P_\alpha-P_0&=\sum_{\textrm{edge }j \ \in \textrm{ cell } \alpha}L_j - P_0\nonumber\\
    &=\sum_{\textrm{edge }j \ \in \textrm{ cell } \alpha}\bigg\lvert\sum_n I_{jn}\mathbf{X}_n\bigg\lvert - P_0.
\end{align}
If we define a cell-edge adjacency matrix $A_{\alpha j}$ by:
\begin{equation}
A_{\alpha j} = \left\{ \begin{array}{ccl}
1, & & \textrm{edge }j \in \textrm{ cell } \alpha\\
0, & & \textrm{otherwise}
\end{array}\right.
\end{equation}
it allows us to rewrite $f_\alpha$ to show the dependence on $\alpha$ more explicitly:
\begin{equation}
    f_\alpha = \sum_{j}A_{\alpha j}\bigg\lvert\sum_n I_{jn}\mathbf{X}_n\bigg\lvert - P_0,
\end{equation}
where $j$ now runs through all the edges and $n$ through all vertices.

Now, we perturb vertex positions $\mathbf{X}_n \to \mathbf{X}_n+\delta \mathbf{X}_n$. We get for $\delta f_\alpha$ {an expression that is} similar to Eq.~(\ref{eq:deltaf_spring}):
\begin{align}
    &\delta f_\alpha= \sum_m \mathbf{R}_{\alpha m} \cdot \delta\mathbf{X}_m+\frac{1}{2}\sum_{j}A_{\alpha j}\bigg[\frac{\left(\sum_n I_{j n}\delta\mathbf{X}_n\right)^2}{L_j}\nonumber\\
    &- \frac{\left(\sum_{nm} I_{j n}I_{j m}\mathbf{X}_n  \cdot \delta\mathbf{X}_m\right)^2}{L_j^3}\bigg].
    \label{eq:deltaf_vertex}
\end{align}
Where we have defined the rigidity matrix as
\begin{equation}
    \mathbf{R}_{\alpha m} = \sum_{j}\sum_{n} \frac{A_{\alpha j}I_{j n}I_{j m}\mathbf{X}_n}{L_j}.
\end{equation}
Note that the last term in Eq.~(\ref{eq:deltaf_vertex}) cannot be written in terms of $\mathbf{R}_{\alpha m}$ anymore.

The self stresses of $\mathbf{R}_{\alpha m}$ impose the following quadratic constraints on zero modes $\delta \mathbf{X}_n$:
\begin{align}
    &\sum_{\alpha} \sigma_{\alpha,I} \sum_j A_{\alpha j}\nonumber\\
    &\bigg[\frac{\left(\sum_n I_{j n}\delta\mathbf{X}_n^{(0)}\right)^2}{L_j} - \frac{\left(\sum_{nm} I_{j n}I_{j m}\mathbf{X}_n  \cdot \delta\mathbf{X}_m^{(0)}\right)^2}{L_j^3}\bigg]=0,
    \label{eq:secondorder-vertex}
\end{align}
which simplifies to
\begin{align}
    \sum_{\alpha} \sigma_{\alpha,I} \sum_j A_{\alpha j}\frac{\left(\delta\mathbf{L}_j^\perp\right)^2}{L_j}=0.
    \label{eq:secondorder-vertex_2-app}
\end{align}

\begin{figure}
    \centering
    \includegraphics[width=0.8\linewidth]{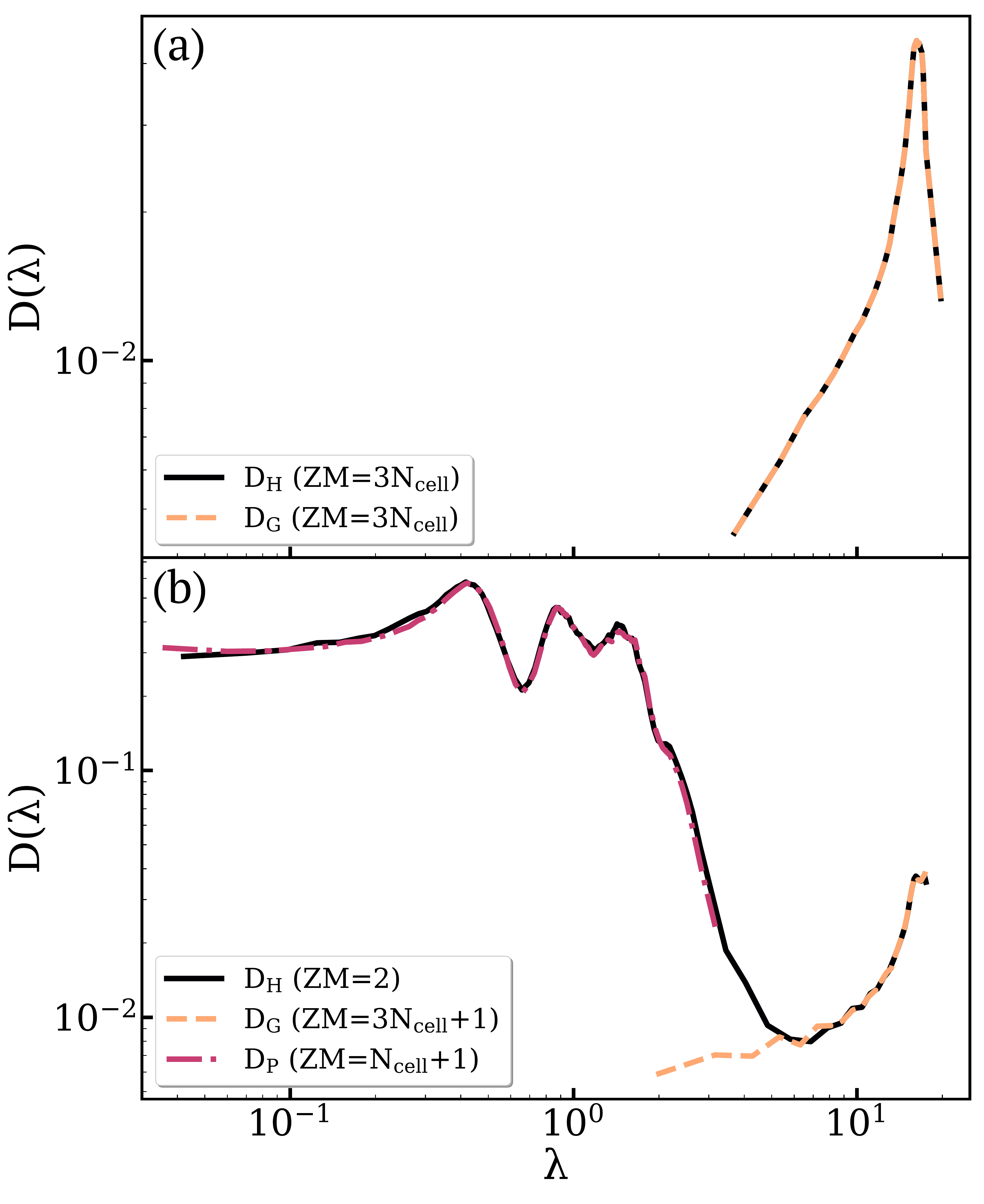}
    \caption{\textbf{Vertex Model ($K_A=0$) Density of States} $D$ as a function of the eigenvalue $\lambda$ for the Hessian matrix (solid black line), Gram matrix (dashed yellow line) and prestress matrix (dash-dotted red line), averaged over 10 samples with $N_{cell}=500$ and and $K_P=1$. (a) In the Floppy regime ($P_0 = 3.90$), the Hessian and the Gram term both possess $3N_{cell}$ zero modes and their DOS curves overlap as there is no prestress in the network. (b) In the rigid regime ($P_0 = 3.70$), the Gram term possesses $3N_{cell}+1$ zero modes and $N_s =1$, while the Hessian only possesses two trivial (translational) zero modes. The Hessian DOS is dominated by the Gram term at high frequency and by the prestress matrix at low frequencies.}
    \label{fig:vertexKA0}
\end{figure}

\begin{figure}
    \includegraphics[width=0.8\linewidth]{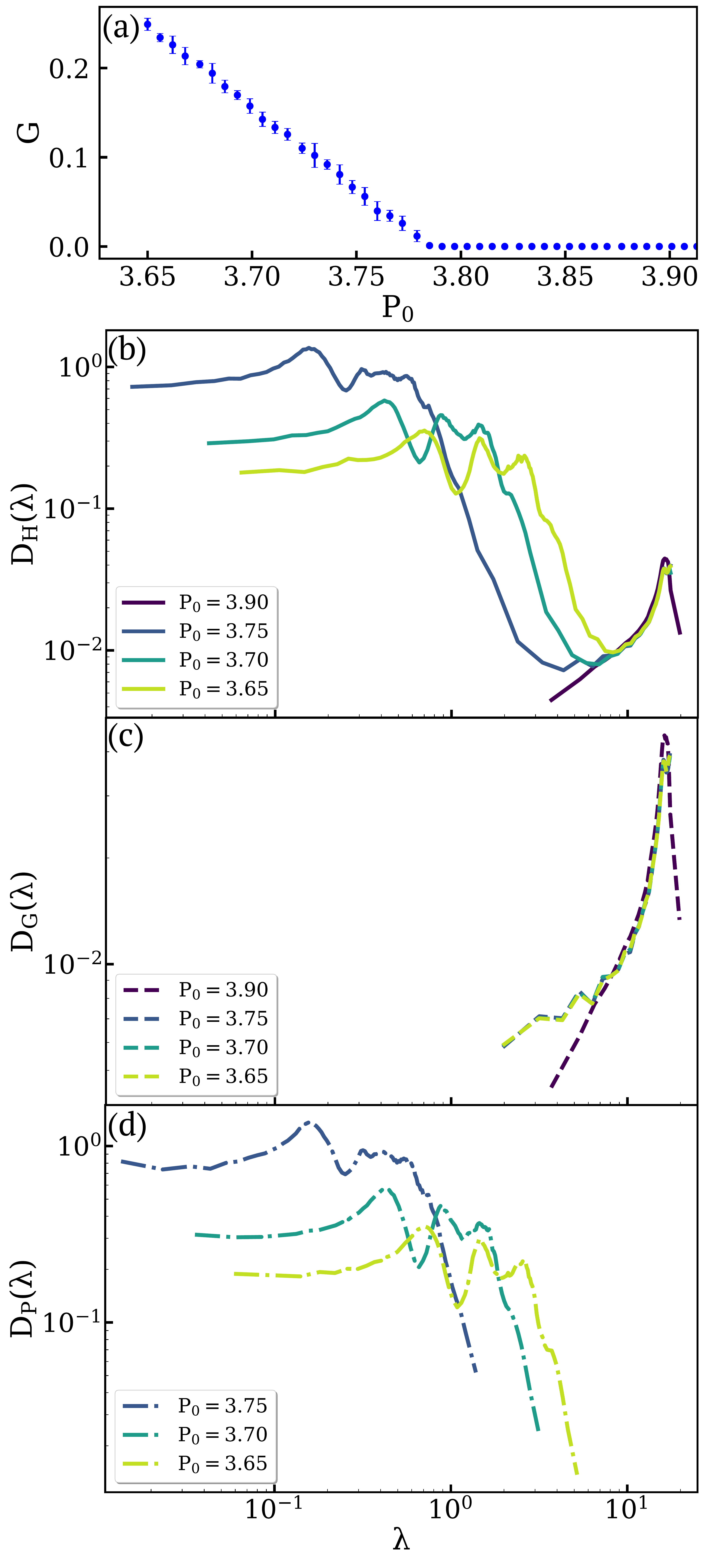}
    \caption{\textbf{Vertex Model ($K_A=0$) Comparison of Density of States} as a function of the eigenvalue $\lambda$ for the Hessian matrix ($D_H$), Gram matrix ($D_G$) and prestress matrix ($D_P$), averaged over 10 samples with $N_{cell}=500$ for different values of $P_0$. (a) Shear modulus ($G$) with as a function of $P_0$ shows a transition from fluid to solid at $P_0^* \approx 3.78$. The error bars represent standard deviation over 10 samples. (b-d) Comparison between DOS of Hessian and its components for different values of $P_0$. (b) Transitioning from floppy to rigid coincides with the appearance of low frequency eigenmodes in the Hessian DOS that shift to higher frequencies as the system becomes more rigid. (c) Even though in the rigid regime there is only one new zero mode, the Gram matrix DOS looks significantly different when compared to systems with $P_0<P_0^*$ because the geometry of the system does not change significantly in the rigid regime. (d) Unlike the Gram matrix, the prestress increases linearly as $L_0$ is decreased in the rigid regime, as reflected by the shift in the prestress matrix DOS towards higher frequencies. Unlike the vertex model with $K_A\neq0$, the prestress matrix has no negative eigenvalues.}
    \label{fig:vertexKA0_compare}
\end{figure}

As discussed in the main text, the inner sum is positive definite. Thus, if a self stress $\sigma_{\alpha,I}>0$ exists, the system is second-order rigid. To see if the shear modulus is non-zero, we again look at the $G=0$ condition (Eq.~(\ref{eq:floppiness-condition})):
\begin{align}
    \sum_{\alpha} f_\alpha \sum_j A_{\alpha j}\frac{\left(\delta\mathbf{L}_j^\perp\right)^2}{L_j}
    = - \sum_\alpha \left( \sum_m \mathbf{R}_{\alpha m} \cdot \delta\mathbf{X}_m\right)^2,
    \label{eq:floppiness-condition-vertex-app}
\end{align}
which cannot be satisfied for any nontrivial zero mode if $f_\alpha >0$. In simulations, we indeed observe that a single self stress exists and $f_\alpha \propto \sigma_\alpha >0$, thus the system is second-order rigid and $G>0$. In Fig.~\ref{fig:vertexKA0} and \ref{fig:vertexKA0_compare} we show plots for vertex model simulations with $K_A=0$ similar to Fig.~\ref{fig:vertex} and \ref{fig:vertex_compare}.  

\subsection{Prestresses in the floppy regime of vertex model with $K_A\neq0$}
It is numerically possible for vertex model configurations in the $P_0>P_0^*$ regime to be prestressed locally. This phenomenon has been reported before~\cite{Merkel2019}. Likewise, we have encountered some cases with four-sided polygons that were prestressed at $P_0=3.90$. {This is because} those four-sided polygons could not achieve both {their preferred area and perimeters, $A_0$ and $P_0$}, even with a zero shear modulus as the prestress is localized. Figs.~\ref{fig:vertex} and \ref{fig:vertex_compare} in the main text exclude such cases.

\section{Numerical methods}\label{app:methods}
\subsection{Structure initialization for spring networks and vertex model}
For both spring networks and vertex model, we use cellGPU~\cite{Sussman2017} to initialize $N_{cell}$ cell centers randomly in a periodic box of size $L_xL_y = N_{cell}$ with $L_x=L_y=\sqrt{N_{cell}}$. A Voronoi tessellation is applied to get $N_{cell}$ polygon cells with $2N_{cell}$ vertices with coordination number $z=3$. The final step in the initialization process involves moving the cell centers for a few time steps using a self-propelled Voronoi model \cite{Bi2016} to make cell areas more uniform. After the initialization process, the energy (Eq.~(\ref{eq:springnet-energy}) for spring networks and Eq.~(\ref{eq:vertexE}) for vertex model) is minimized by moving the vertices using the FIRE minimizer \cite{Bitzek2006} with a force cutoff of $10^{-12}$. For vertex model, a $T1$ transition was performed when an edge length became smaller than $0.01$. The size of the time steps for the simulations were dynamically decided by the minimizer, starting from $\mathrm{d}t=0.001$, but allowed to be increased up to $\mathrm{d}t_{max}=0.1$.

\subsection{Structure initialization for jammed packings}
We create 2D disk packings using a quad-precision GPU implementation of the FIRE algorithm~\cite{Charbonneau2012,Morse2014}. First, $N$ particles with a polydispersity of $20 \%$ are randomly distributed in a periodic box of size $L = 1$ and then radii are re-scaled to a packing density well above jamming transition which is typically $\phi_J \simeq 0.84$ for 2D systems. Finally, the system is minimized to its inherent structure using the FIRE minimizer with a force cutoff of $10^{-20}$. At densities far from jamming, a packing will have many states of self stress. To bring the system to the critical jamming with one state of self stress, we successively re-scale the density to smaller values and re-minimize the energy. Once the system reaches one state of self stress, the pressure will be in order of $p \approx 10^{-8}$ and the initialization process is halted.

\subsection{Density of states and shear modulus}
To find the eigenvalue spectrum of the Hessian, Gram term and prestress matrix, we calculate the Hessian matrix and its components for a given system at an energy minimum and consider eigenvalues with an absolute value smaller than $10^{-10}$ as zero eigenmodes. For the density of states plot, we sort all the eigenvalues and use equiprobable (Dirichlet) binning with 150 bins such that there is an equal number of eigenvalues in each bin, from which we can plot a normalized histogram representing the density of states. We also apply a centered moving average with window size 3 to smooth the curves. Zero modes would represent a peak at 0 which are not plotted. For the shear modulus plots, we modify the periodic boundary conditions to accommodate a skew (i.e.\ Lees-Edwards boundary conditions) with a simple shear parameter $\gamma$, which allows us to write the energy function as a function of $\gamma$ \cite{Merkel2018}. We then use Eq.~(\ref{eq:shearModulus}) to calculate the shear modulus. 

\subsection{Numerical results for vertex models with no area constraints ($K_A = 0$)}

Using the same analyses as discussed for Figs.~\ref{fig:vertex} and \ref{fig:vertex_compare}, we calculate the contributions to the Hessian density of vibrational states for vertex models where there are no area constraints, ($K_A = 0$). These data are shown in Fig~\ref{fig:vertexKA0} and~\ref{fig:vertexKA0_compare}, respectively.

\bibliography{apssamp}

\end{document}